# Anisotropy-Induced Photonic Bound States in the Continuum

**Authors:** Jordi Gomis-Bresco[1], David Artigas[1,2] and Lluis Torner[1,2]*

**Affiliations:**

[1] ICFO-Institut de Ciencies Fotoniques, The Barcelona Institute of Science and Technology, Av. Carl Friedrich Gauss, 3, 08860 Castelldefels (Barcelona), Spain.

[2] Department of Signal Theory and Communications, Universitat Politècnica de Catalunya, Jordi Girona 1-3, 08034 Barcelona, Spain.

*Correspondence to: lluis.torner@icfo.eu

**Bound states in the continuum (BICs) are radiationless localized states embedded in the part of the parameter space that otherwise corresponds to radiative modes. Many decades after their original prediction[1,2] and early observations in acoustic systems[3], such states have been recently demonstrated in photonic structures with engineered geometries[4–17]. In this paper we put forward a mechanism, based on waveguiding structures containing anisotropic birefringent materials, which affords the existence of BICs that exhibit fundamentally-new properties. In particular, anisotropy-induced BICs may exist in symmetric as well as in asymmetric geometries; they form in tunable angular propagation directions; their polarization may be pure transverse-electric, pure transverse-magnetic or full-vector with tunable polarization-hybridity; and they may be the only possible bound states of properly designed structures, thus appearing as a discrete, isolated bound state embedded in a whole sea of radiative states.**

Bound states in the continuum were originally predicted in 1929 by von Neumann and Wigner as discrete fully-bounded quantum states with energies above the continuum of the corresponding Hamiltonian. Signatures of acoustic BICs were observed experimentally several decades ago[3], but only after the recent theoretical studies and landmark experiments conducted in classical photonic systems the implications of the almost century-old concept were properly appreciated, stimulating deeper understanding over its origin[9–11] and inspiring new schemes[12–14] and applications[15,16]. BICs discovered to date are almost-pure transverse-electric (TE) or transverse-magnetic (TM) waves[4,7–9,18–20], namely with a negligible fraction of energy being carried by the respective orthogonal polarization. The corresponding trapping mechanism can thus be intuitively viewed as a scalar or spinor potential. In contrast, photonic structures containing anisotropic media can support bound states that, in general, involve the full-vector electric and magnetic field components. This important feature opens up the possibility to search for full-vector BICs that cannot exist in scalar analogies. In this paper we explore the concept and expose its potential.

By and large, coupling localized states with a coexisting radiative continuum results in energy being shed away and thus in unbound states that decay and fade away during propagation. Bound states can only exist when such transverse radiative leakage is suppressed by a suitable mechanism. Here we address the existence of such mechanism in optical waveguiding structures fabricated in anisotropic media. We restrict ourselves to the simplest structures fabricated in birefringent uniaxial natural materials, the optical axes of the crystals are set to lie on the waveguide plane forming an angle $\theta$ with the wave propagation direction, $\theta$ is set to be equal for all materials, and we set the cladding to be an isotropic medium (Fig. 1a). However, the concept is general and holds for a wealth of more complex structures, including multilayer geometries[21], off-plane and unaligned optical axes, and structures containing different anisotropic media, including biaxial media and other types of natural or artificial anisotropic materials[22–24].

In birefringent media, for an arbitrary direction of the light propagation relative to the optical axis, light fields are a superposition of ordinary and extraordinary waves, which are respectively characterized by the ordinary refractive index $n_o$ and the angle-dependent refractive index $n_e(\theta)$ that varies between $n_o$ and the extraordinary refractive index $n_e$, as $n_e^2(\theta) = n_e^2 n_o^2 /(n_e^2\cos^2\theta + n_o^2\sin^2\theta)$. All $n_{os}$, $n_{of}$, $n_{es}$ and $n_{ef}$, where the subscripts "s, f" stand for the substrate and the core layer, respectively, are assumed to be larger than the refractive index of the cladding material, which for simplicity here we set to be air ($n_c=1$). The light cone is therefore defined as $\omega=(c\cdot k_0)/n$ where $\omega$ and $c$ are the angular frequency and the vacuum speed of light, respectively, $k_0 = 2\pi/\lambda_0$ is the vacuum wavenumber, $\lambda_0$ is the vacuum wavelength, and $n = max\,(n_{es}(\theta)$ or $n_{os})$.

Bound states in structures containing such media are characterized by the effective index $N=\beta/k_0$, where $\beta$ is the propagation constant and, in general, they involve the full-vector six field components and exhibit two different decay constants, $\gamma_o$ and $\gamma_e$, for the transversally-evanescent ordinary and extraordinary fields in the substrate. When $N > n$, propagation occurs below the light cone as standard guided states, with real $\gamma_o$ and $\gamma_e$. Above the light cone, $N < n$ and therefore $\gamma_o$ ($\gamma_e$) becomes an imaginary quantity if the substrate is a birefringent negative (positive) material while $\gamma_e$ ($\gamma_o$) remains real. Then the ordinary (extraordinary) wave does not describe an evanescent field but a radiative wave. Under such conditions, partially localized states may still exist in the form of leaky[25–27] or, more properly, semi-leaky waves. Such waves are partially-localized improper modes, in the sense that they carry an infinite amount of energy in the region far from the waveguide, but they may be used to describe non-stationary states that leak energy that is carried away towards the substrate by a *continuous band of radiation modes.* For small radiative losses, the field in the vicinity of the core—substrate interface corresponding to such continuous band can be approximately described by a single leaky mode that features a complex $N$ whose imaginary part gives a good approximation to the actual radiation losses. In contrast to isotropic structures, which have a single radiation channel, by their very nature such leaky modes exhibit two possible radiation channels corresponding to the ordinary and the extraordinary waves, but only one of them may radiate. This property is afforded by the fully-vectorial nature of the states and provides a powerful way to control their leakage losses. In particular, BICs may form when the radiation channel of the leaky mode is suppressed, either because the corresponding ordinary or extraordinary radiating wave is not required for a bound state to exist, or, more importantly, because of a destructive interference of the radiative waves. Both mechanisms can be induced by anisotropy, as shown in Fig. 1b. The plot displays the frequency-momentum diagram of an illustrative structure made of a positive birefringent substrate and a negative birefringent core layer film. The structure supports two families of standard guided modes, in the areas shown as blue surfaces below the light cone, and leaky modes, in the areas depicted as orange surfaces above the light cone. The corresponding dispersion relations as a function of the optical axis orientation are shown in Fig. 1 of the supplementary material. The radiative channel is the extraordinary wave and the light cone is defined by the extraordinary refractive index. The decay length of the leaky modes, defined as the inverse of the imaginary part of $N$, is shown in Fig. 1c for a structure with $d/\lambda=0.22$. The decay length exhibits discrete diverging peaks at specific off-axis optical axis orientations, revealing the existence of fully bound states embedded in the leaky branch.

The central insight put forward in this paper is that the interplay between the values of the ordinary and extraordinary refractive indices of the core layer and the substrate, the positive or negative nature of the crystals, the optical axes orientations, the cladding refractive index, and the waveguide thickness for a given operating wavelength/frequency, afford the existence of BICs with pure-TE, pure-TM and tunable full-vector polarization, and in a wealth of modal spectroscopies. Some of the different possibilities are illustrated in Figure 2, which displays isofrequency sections of the frequency-momentum diagram obtained for six different suitable material configurations.

In all plots, the BICs that occur at discrete points embedded in the leaky branches are displayed as blue dots. Figure 2a illustrates the existence of BICs with properties similar to those previously reported in photonic crystals[8–10]. Namely, a TM-polarized BIC that co-exists with standard guided modes elsewhere in the parameter space. The BIC occurs as a bounded state propagating *exactly* at 90º relative to the crystal optical axis, surrounded by radiating modes in all angular (polar and azimuthal) directions. The refractive indexes in the structure fulfil $n_c=1 < n_{os} < n_{es} < n_f$. Such BICs cease to exist in the very same structure if, e.g., the birefringence strength of the substrate material is reduced, as shown in Fig. 2b. A fundamentally different modal spectroscopy occurs when the index of the isotropic core layer is set between the extraordinary and the ordinary indexes of a positive birefringent material used as a substrate (i.e., $n_c=1 < n_{os} < n_f < n_{es}$). Then, guided modes propagating perpendicularly to the optical axis are not possible at all, but a discrete TM-polarized BIC does still exist at exactly such orientation, as shown by Fig. 2c. Structures that support on-axis as well as off-axis BICs can be obtained by turning the core layer birefringent, as illustrated in Fig. 2d (where $n_c=1 < n_{os} < n_{ef} < n_{of} < n_{es}$). Finally, Fig. 2e (where $n_c=1 < n_{es} < n_f < n_{os}$) and Fig. 2f (where $n_c=1 < n_{es} < n_{of} < n_{ef} < n_{os}$) correspond to structures build in a negative birefringent material as a substrate with core layer having refractive indexes laying between the extraordinary and ordinary indexes of the substrate. Under such conditions, standard guided modes are not allowed anywhere, but discrete BICs do exist, therefore being the only fully-bound states supported by the structure. They occur as TE-polarized states or as full-vector states. Other modal spectroscopies (not shown in the plot), such as multiple TE-polarized or TM-polarized BICs only or full-vector BICs only (the latter case requires relaxing the condition $\theta_s = \theta_f$), are also possible.

The physical origin and nature of the different BICs is critically dictated by the light propagation direction relative to the crystal optical axis. When light propagates orthogonally to the optical axis, the modes are polarization-separable. Then BICs decouple from the continuum because they are pure-TE (pure-TM) for negative (positive) birefringent media, and thus orthogonal to that of the radiation channel (see supplementary Fig 2). Such BICs can occur near the cut-off of properly designed standard integrated optical waveguides, such as those fabricated by proton-exchange, and, in particular, realize and generalize the concept of critically stable modes[26]. In contrast, for an arbitrary propagation direction, the mechanism that makes full-vector BICs possible is the total destructive interference of the radiation fields. Such an unusual phenomenon may occur when the properties of the structure are prepared to allow the existence of special orientations at which the radiative fields at the interface between the core layer and the substrate exactly vanish. Imposing this condition in the general eigenvalue equation for bound states leads to Eqs. (9)-(10) of the supplementary material, which thus provides the algorithm to search for full-vector interferometric BICs. Suitable conditions occur, e.g., when the core layer is a birefringent material and thus a phase difference between the ordinary and extraordinary field components that makes the destructive interference possible is introduced, as shown in Fig. 2d and Fig. 2f. Because of its physical roots in a null-interference point, the existence condition Eqs. (9)-(10) is adiabatically robust against variations in the structure, hence the robustness of the corresponding BICs (see supplementary Fig. 3). Note that full-vector BICs are the states that realize the radiative loss minima of the leaky modes propagating in the symmetric waveguides studied in ref [28]. However, importantly, it must be properly appreciated that the existence of full-vector BICs is a general phenomenon, not at all restricted to symmetric geometries, as shown here. We also verified that they exist in highly asymmetric geometries, as well as in multi-layer, multi-material and, importantly, high-contrast ultrathin structures.

Full-vector interferometric BICs exhibit a set of unique properties. For example, importantly, they are families of states following a line in the 3D frequency-momentum diagram (Fig. 1a), in sharp contrast to usual BICs[8,9,29] that exist as a single point in that representation. Also, in contrast to previously reported TE-like or TM-like BICs, the full-vector nature of the interferometric BICs allows tuning their TE / TM polarization content to a large degree, by varying structural or operational control parameters, such as the core layer thickness or the wavelength. This is shown in Fig. 2g for the relevant BIC displayed in Fig 2d. Also, in contrast to standard BICs existing in photonic crystals, which occur only along symmetry-axes, the angular loci at which the full-vector interferometric anisotropy-induced BICs occur can be readily tuned. This is illustrated in Fig. 2h and 2g for the BIC displayed in Fig 2d. Note that the TE/TM polarization content of the BIC also varies along the curve in Fig 2h, consistent with Fig 2g, showing that various external parameters can be used to tune the practical properties of the full-vector BICs. A possibility of particular practical interest is illustrated in Fig 2i, which illustrates how the BIC propagation direction can actively be controlled by changing any of the refractive indices of the structure. To highlight the point, the figure explores a variation of the refractive index over a large range ($\Delta n = 0.2$), which causes a change of the BIC existence angular loci of more than $\Delta\theta \sim 40°$. Note that $\Delta n = 0.2$ is however compatible with the change of the extraordinary refractive index of a liquid crystal under temperature tuning[30]. Additionally, Supplementary Fig. 4 shows the variation of the angular propagation direction of BICs generated in a suitable (yet, non-optimized) proton exchange $LiNbO_3$ waveguide structure under the action of the Pockels effect. The plot illustrates that an applied external field of the order of 1 V/μm may result in a variation of the BIC angular existence loci of more than ten degrees. Such large angular loci steering, as well as the accompanying sharp transitions from bound modes to radiating states, may find applications in integrated photonic devices for sensing, spatial-light modulation or filtering[31,32].

Figure 3 shows the experimental signature of the anisotropy-induced BICs, obtained by modal spectroscopy spectra in a prism-coupling geometry (see methods, and supplementary Fig. 5). We used calcite as the substrate material and a spin-coated polymer for the core layer film to form an anti-guiding configuration that features the salient phenomenon described in Fig. 2e, where BICs are the only existing non-radiating bound states. The plots compare the theoretical and the experimental reflectance when TE-polarized light is used as illumination. Figure 3a shows that above the light cone the incoming light is fully reflected except when it couples to a radiative leaky mode or to a TE-polarized BIC, resulting then in a dip in the image. Because of the hybrid polarization of the leaky modes, light coupled to the modes is partially converted to TM-polarization, which in this structure is predominantly an ordinary wave. Therefore, the TE-TM polarization conversion reflectance shown in Fig. 3b (see also the corresponding transmittance in supplementary Fig. 6) is an indirect indication of the leakage losses of the different states. At light propagation orthogonal to the crystal optical axis ($\theta = 90º$), the two branches in Fig. 3b (see also zooms in 3c and 3d) exhibit no visible polarization conversion reflection, consistent with the existence there of the two fully bound states displayed in Fig. 2e. Note that, in contrast to previous experiments done in photonic crystals[8], our sample is highly asymmetric and BICs are excited by prism coupling from the cladding and not from the substrate, thus Fano resonances do not occur.

In closing, we stress that the occurrence of BICs in fully-vectorial settings opens the door to the exploration of such bound states beyond traditional scalar or spinor analogies. Anisotropy affords a new radiation suppression mechanisms based on the concept of semi-leaky modes by which BICs readily can occur in both, symmetric and highly asymmetric structures. All modal spectroscopy combinations, i.e., modes-only, modes-and-BICs, and BICs-only, may be realized. Note that in the latter case, BICs are not higher-order modes but rather the only possible bound state. Anisotropy-induced BICs can be polarization-separable or intrinsically polarization-hybrid. Full-vector BICs exhibit tunable angular propagation direction and tunable polarization-hybridity, as well as highly-directional, ultra-sharp transitions from radiationless to radiative states, properties that may find applications in photonic filters, spatial-light modulators and sensors based on angular selectivity. We anticipate similar phenomena in off-plane geometries, multilayer and multi-material structures, biaxial crystals and generalized anisotropic media, such as chiral and hyperbolic materials as well as engineered natural and artificial materials crafted in geometries, including high-contrast ultrathin structures and metasurfaces, with form-anisotropy.

**Methods**

**Theoretical and Numerical Methods:** A detailed description of the derivation of the eigenvalue equations for interferometric BICs and the related transfer matrix and beam propagation methods are provided as Supplementary Information.

**Experimental Methods:** Modal spectroscopy spectra in a prism-coupling geometry were performed to demonstrate the BIC existence. The experimental layout is shown in Supplementary Fig. 6. The spin-coated polymeric film used as a core film was measured using profilometry and ellipsometry techniques, resulting in a thickness $d$ = 1.6 μm and $n_f$ = 1.553+0.0005i. We placed a 5-mm equilateral SF11 prism (n =1.779) on top of the sample, using a liquid to match the prism refractive index and minimize any air gap between the sample and the prism. We focused an 8 mm diameter laser beam ($\lambda_0$ = 632.8 *nm*) with an $f$ = 25 mm lens on the top surface of the sample through the tilted prism facet. The light reflected by the sample, coming out from the opposite prism facet, illuminated a CCD camera with its sensing surface (6 mm wide x 8 mm high) perpendicular to the reflection direction, in such a way that the vertical axis of the CCD image provided in one shot the dependence with the angle of incidence ($\varphi$). The incident and reflected light polarizations were selected by means of two independent polarizers. The propagation direction (angle θ with respect to the sample's OA) was defined by the plane of incidence and controlled by rotating the sample with a goniometer. The measurement was performed changing the propagation angle in steps of 2 degrees and building Figure 3 by stitching together the central part of the images obtained.

**Acknowledgements:** The authors thank Paula Mantilla for the spin coating of the calcite samples and Rob J. Sewell for discussions in writing the paper. This work has been partially supported by Fundation Cellex, Fundation Mir-Puig and the Spanish government through grant FIS2015-71559-P and the Severo Ochoa Excellence Program.

**Author Contributions**: All authors contributed equally to the work.

**Competing financial interests:** The authors declare no competing financial interests.

**Bibliography:**

1. Neumann, J. von & Wigner, E. P. Über merkwürdige diskrete Eigenwerte. *Phys. Z.* **30,** 465–467 (1929).

2. Stillinger, F. H. & Herrick, D. R. Bound states in the continuum. *Phys. Rev. A* **11,** 446–454 (1975).

3. Parker, R. Resonance effects in wake shedding from parallel plates: Some experimental observations. *J. Sound Vib.* **4,** 62–72 (1966).

4. Marinica, D. C., Borisov, A. G. & Shabanov, S. V. Bound states in the continuum in photonics. *Phys. Rev. Lett.* **100,** 1–4 (2008).

5. Bulgakov, E. N. & Sadreev, A. F. Bound states in the continuum in photonic waveguides inspired by defects. *Phys. Rev. B - Condens. Matter Mater. Phys.* **78,** 1–8 (2008).

6. Moiseyev, N. Suppression of feshbach resonance widths in two-dimensional waveguides and quantum dots: A lower bound for the number of bound states in the continuum. *Phys. Rev. Lett.* **102,** 1–4 (2009).

7. Plotnik, Y. *et al.* Experimental observation of optical bound states in the continuum. *Phys. Rev. Lett.* **107,** 28–31 (2011).

8. Hsu, C. W. *et al.* Observation of trapped light within the radiation continuum. *Nature* **499,** 188–91 (2013).

9. Zhen, B., Hsu, C. W., Lu, L., Stone, A. D. & Soljačić, M. Topological Nature of Optical Bound States in the Continuum. *Phys. Rev. Lett.* **113,** 1–5 (2014).

10. Yang, Y., Peng, C., Liang, Y., Li, Z. & Noda, S. Analytical perspective for bound states in the continuum in photonic crystal slabs. *Phys. Rev. Lett.* **113,** 1–5 (2014).

11. Hsu, C. W., Zhen, B., Stone, A. D., Joannopoulos, J. D. & Soljačić, M. Bound states in the continuum. *Nat. Rev. Mater.* **1,** 16048 (2016).

12. Monticone, F. & Alù, A. Embedded photonic eigenvalues in 3D nanostructures. *Phys. Rev. Lett.* **112,** 1–5 (2014).

13. Gao, X. *et al.* Formation mechanism of guided resonances and bound states in the continuum in photonic crystal slabs. *Sci. Rep.* **6,** 31908 (2016).

14. Rivera, N. *et al.* Controlling Directionality and Dimensionality of Wave Propagation through Separable Bound States in the Continuum. *Sci. Rep.* 33394 (2016).

15. Lepetit, T. & Kanté, B. Controlling multipolar radiation with symmetries for electromagnetic bound states in the continuum. *Phys. Rev. B* **90,** 1–4 (2014).

16. Dreisow, F. *et al.* Adiabatic transfer of light via a continuum in optical waveguides. *Opt. Lett.* **34,** 2405–2407 (2009).

17. Lee, J. *et al.* Observation and differentiation of unique high-Q optical resonances near zero wave vector in macroscopic photonic crystal slabs. *Phys. Rev. Lett.* **109,** (2012).

18. Molina, M. I., Miroshnichenko, A. E. & Kivshar, Y. S. Surface bound states in the continuum. *Phys. Rev. Lett.* **108,** 1–4 (2012).

19. Corrielli, G., Della Valle, G., Crespi, A., Osellame, R. & Longhi, S. Observation of surface states with algebraic localization. *Phys. Rev. Lett.* **111,** 1–5 (2013).

20. Weimann, S. *et al.* Compact surface fano states embedded in the continuum of waveguide arrays. *Phys. Rev. Lett.* **111,** 1–5 (2013).

21. Shipman, S. P. & Welters, A. T. Resonant electromagnetic scattering in anisotropic layered media. *J. Math. Phys.* **54,** (2013).

22. Krishnamoorthy, H. N. S., Jacob, Z., Narimanov, E., Kretzschmar, I. & Menon, V. M. Topological transitions in metamaterials. *Science* **336,** 205–209 (2012).

23. Jahani, S. & Jacob, Z. All-dielectric metamaterials. *Nat. Nanotechnol.* **11,** 23–36 (2016).

24. Poddubny, A., Iorsh, I., Belov, P. & Kivshar, Y. Hyperbolic metamaterials. *Nat. Photonics* **8,** 78–78 (2013).

25. Monticone, F. & Alú, A. Leaky-Wave Theory, Techniques, and Applications: From Microwaves to Visible Frequencies. *Proc. IEEE* **103,** 793–821 (2015).

26. Knoesen, A., Gaylord, T. K. & Moharam, M. G. Hybrid guided modes in uniaxial dielectric planar waveguides. *J. Light. Technol.* **6,** 1083–1104 (1988).

27. Torner, L., Recolons, J. & Torres, J. P. Guided-to-leaky mode transition in uniaxial optical slab waveguides. *J. Light. Technol.* **11,** 1592–1600 (1993).

28. Marcuse, D. & Kaminow, I. Modes of a symmetric slab optical waveguide in birefringent media, part II: Slab with coplanar optical axis. *IEEE J. Quantum Electron.* **15,** 92–101 (1979).

29. Hsu, C. W. *et al.* Bloch surface eigenstates within the radiation continuum. *Light Sci. Appl.* **2,** e84 (2013).

30. Li, J., Gauza, S. & Wu, S.-T. Temperature effect on liquid crystal refractive indices. *J. Appl. Phys.* **96,** 19 (2004).

31. Gomez-Diaz, J. S. & Alu, A. Flatland Optics with Hyperbolic Metasurfaces. *ACS Photonics* acsphotonics.6b00645 (2016). doi:10.1021/acsphotonics.6b00645

32. Smalley, D. E., Smithwick, Q. Y. J., Bove, V. M., Barabas, J. & Jolly, S. Anisotropic leaky-mode modulator for holographic video displays. *Nature* **498,** 313–7 (2013).

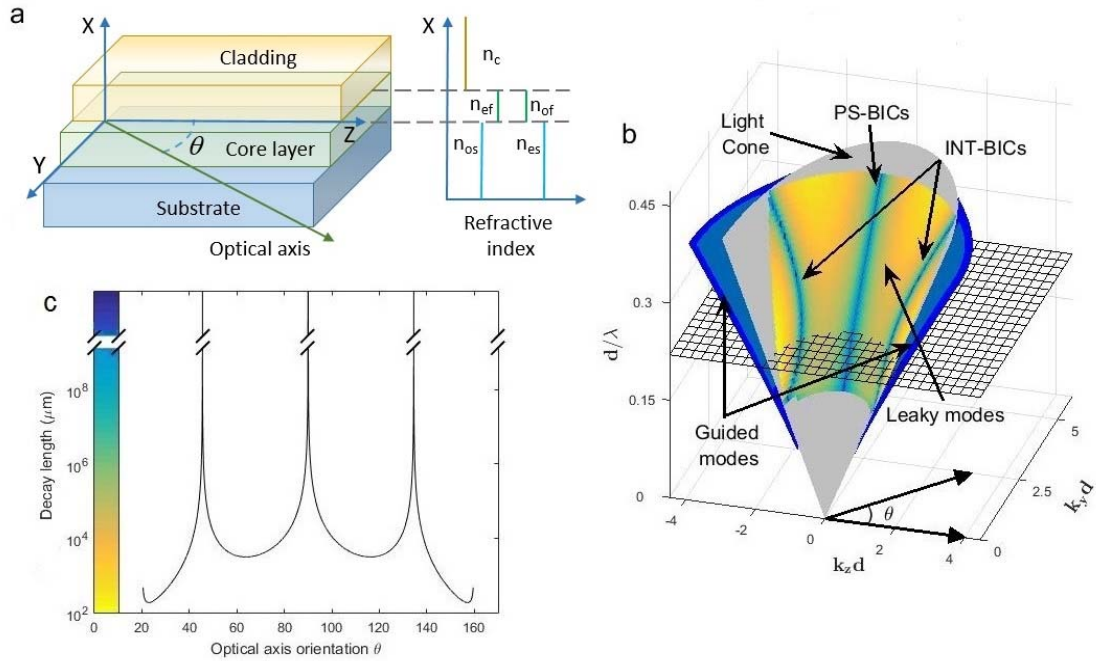

**Fig. 1**. **Modal spectroscopy. a,** Schematic of a waveguiding structure fabricated with birefringent materials for the core layer and substrate. A typical relation among the various refractive indexes is shown in the inset: The higher refractive index in the structure is featured by either the ordinary ($n_{os}$) or the extraordinary ($n_{es}$) index in the substrate, in general the ordinary ($n_{of}$) and extraordinary ($n_{ef}$) indexes of the core lay between those of the substrate, and in all the cases analyzed here the cladding is chosen to be air ($n_c$ =1). **b,** Light cone (grey surface) for a structure composed by a negative-birefringent core ($n_{of}$ = 1.75, $n_{ef}$ = 1.5) on a positive-birefringent substrate ($n_{os}$ = 1.25, $n_{es}$ = 2). Standard guided modes exist in the blue areas and semi-leaky modes exist in the orange areas. Polarization-separable (PS-BICs) and full-vector interferometric (INT-BICs) BICs exist in the loci indicated with blue lines within the yellow area. **c,** Decay length of the leaky states existing at $d/\lambda$=0.22 versus propagation direction relative to the crystal optical axis orientation. This case corresponds to the isofrequency cut indicated in **(b)** by the squared mesh. The peaks that diverge to infinite reveal the existence of discrete BICs at specific orientations. The color bar in **(c)**, used also in **(b),** displays the radiative losses in logarithmic scale calculated using the imaginary part of the effective index of the leaky modes**.**

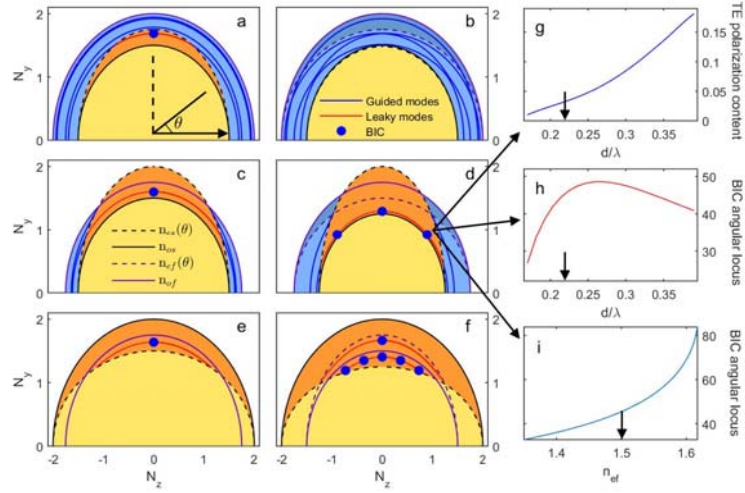

**Fig. 2. Typology of anisotropy-induced BICs.** Isofrequency sections of the light cone diagram for waveguiding structures with different constitutive parameters. Regions below and above the light cone are shown in blue and yellow, respectively. Standard mode branches are shown as blue lines within the blue region. Radiating semi-leaky branches exist in the orange region and are shown as red lines. BICs are the discrete points highlighted with blue dots on the red semi-leaky branch. **a,** TM-polarized BICs coexist with guided modes for an isotropic film ($n_f$=2 and thickness $d$=0.8λ) and a positive-birefringent substrate ($n_{os}$=1.5 and $n_{es}$=1.75). **b,** BICs disapear for substrates with smaller birefringence ($n_{os}$=1.5, $n_{es}$=1.48) and core layers with higher refractive index even if the core layer is anisotropic ($n_{of}$=2, $n_{ef}$=1.75 and $d$=0.8λ). **c,** Guided modes and TM-polarized BICs coexist for an isotropic core layer with refractive index ($n_f$=1.75) between those of the positive-birefringent substrate ($n_{os}$=1.5, $n_{es}$=2) **d,** TE-polarized and multiple full-vector BICs appear for a core layer that is birefringent ($n_{of}$=1.75, $n_{ef}$=1.5, $d$=0.22λ) on a birefringent substrate ($n_{os}$=1.25, $n_{es}$=2). **e, f,** Structures where BICs are the only non-radiating state: **(e)** a TE-polarized BIC in an isotropic core layer ($n_f$=1.75, $d$=0.5λ) on a negative-birefringent substrate ($n_{os}$=2, $n_{es}$=1.5) or **(f)** TE-polarized and multiple full-vector BICs in a structure with a positive-birefringent core layer ($n_{of}$=1.5, $n_{ef}$=1.75, d=0.68λ) on a negative-birefringent substrate ($n_{os}$=2, $n_{es}$=1.25). **g,** Degree of polarization-hibrity of the BICs and **h,** angular loci at which the full-vector interferometric BICs occur, in both cases as a function of the normalized thickness of the core layer. **i,** Angular loci at which the full-vector interferometric BICs occur as a function of the extraordinary refractive index of the film. All **(g)**, **(h)** and **(i)** correspond for the BIC shown in **(d)**. The degree of polarization-hybridity is measured as the fraction of the total BIC's power density carried by the field components correspoding to the TE-polarization. Arrows are an eye-guide pointing to the original case in panel **(d)**.

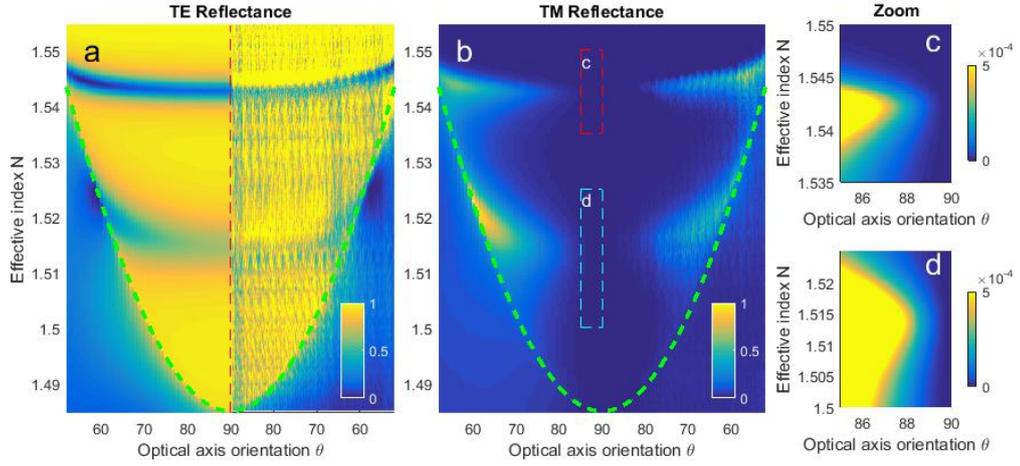

**Fig. 3. Theoretical and experimental modal spectroscopy spectra for an anti-guiding waveguide on a calcite substrate. a**, TE-incident TE-reflected intensity; **b,** TE-incident, TM-reflected intensity. Taking advantage of the specular symmetry of the structure, we compare transfer matrix theoretical calculations (left-hand-side, θ = 0º- 90º) with the experimental observations (right-hand-side, θ = 90º-0º). The green dashed line shows the angular dependence of the extraordinary refractive index for calcite. **c** and **d,** depict theoretical zooms near the region of existence of polarization separable BICs within the dashed-line boxes in **(b)**.